\documentstyle[prd,aps,psfig,floats,twocolumn]{revtex}

\begin{document}
\title {Investigation of a Toy Model for Frustration in Abelian 
Lattice Gauge Theory}

\author{V. Azcoiti}
\address{
Departamento de F\'\i sica Te\'orica \\ 
Facultad de Ciencias, Universidad de Zaragoza\\
50009 Zaragoza (Spain).}

\author{G. Di Carlo}
\address{
Istituto Nazionale di Fisica Nucleare,
Laboratori Nazionali di Frascati \\
P.O.B. 13 - Frascati 00044 (Italy). }

\author{E.~Follana}
\address{
Departamento de F\'\i sica Te\'orica \\ 
Facultad de Ciencias, Universidad de Zaragoza\\
50009 Zaragoza (Spain).}

\date{\today}
\maketitle

\begin{abstract}
We introduce a lattice model with local U(1) gauge symmetry which incorporates
explicit frustration in $d>2$. The form of the action is inspired from
the loop expansion of the fermionic determinant in standard
lattice QED. We study through numerical simulations the phase
diagram of the model, revealing the existence of a frustrated 
(antiferromagnetic) phase for $d=3$ and $d=4$, once an appropriate 
order parameter is identified.
\end{abstract}


\section{Introduction}

The importance of frustration and disorder is well known to condensed matter
physicists working in the field of spin-glasses and related systems 
\cite{Parisi}. In these systems one can find a variety of unusual phases 
which, in some cases, present completely new phenomena as, for instance, 
the existence of a multiplicity of vacua not related by a symmetry of the 
action. It is interesting to ask whether similar phenomena could happen
in a field theory and what will be the physical consequences.

We address here the issue of frustration in lattice gauge theories.
The motivation for this work relies on the fact that, as well known, 
realistic gauge theories with dynamical fermions, in contrast to 
gauge theories with fundamental scalar fields, are characterized 
by a frustrated effective action. Frustration here must be understood 
in the most naive way, i.e., after the integration of the Grassmann 
degrees of freedom, the effective fermion action has contributions 
which can be seen as competing interactions.

Even if spin-glass like phases in gauge theories have been found 
time ago we are interested, as stated before, in more 
realistic gauge theories. More precisely, strongly coupled QED in 
the noncompact lattice regularization has received considerable 
attention in recent time (\cite{Kocic,Azcoiti} and references therein). 
Many interesting and exciting new results in the field of four-dimensional 
gauge theories, as for instance the possibility to have a non Gaussian fixed 
point \cite{Azcoiti}, appear in this model. The origin of these new phenomena 
is unfortunately not clear at all. It has been argued in the past that the 
monopole percolation transition could drive the chiral transition of 
strongly coupled QED \cite{MONO}. This issue is however rather unclear and 
it is therefore worthwhile to explore other possibilities. There are 
furthermore other exciting results in non compact QED as the existence 
of a non chiral phase transition in the three-dimensional model 
\cite{2+1} or the fact that the transition seems to prolongate 
to non vanishing fermion masses \cite{NUESTROIJMP} which allow one to 
speculate that frustration could be the origin of these phenomena. 
We should say from the beginning that at the moment this is only speculation. 
Our aim is to investigate this issue and, as a first step in this direction, 
we will study in this paper a model \cite{Edimburgo}
in which frustration is introduced by hand, although 
inspiration is taken from the hopping-parameter expansion of QED.
Section \ref{frustration} is devoted to review how frustration appears in 
lattice QED. We analyze this phenomenon from the hopping parameter expansion 
as well as from the loop expansion of the determinant of the 
Dirac-Kogut-Susskind operator. In Sec.\ \ref{toy-model} we define our 
toy model, discuss its classical structure of ground states, introduce an 
order parameter for the phase transition and show how our model in two 
dimensions is equivalent to the XY-spin model. Section \ref{numeric} contains 
the numerical results of a simulation and the phase structure. The conclusions
are reported in Sec.\ \ref{comments}.

\section{Frustration in lattice QED}
\label{frustration}

We start with the action functional for QED on the lattice 
using Kogut-Susskind fermions, $S=\beta S_G + S_F $, with $S_G$
the pure gauge action (either compact or noncompact) and $S_F$ the fermionic 
action,

\begin{eqnarray}
S_F= m\sum_x \overline\chi(x) \chi(x)+ \nonumber \\
 \frac{1}{2}\sum_{x,\mu} \eta_\mu(x)\overline\chi(x)\{U_\mu(x)\chi(x+\mu)-
U_\mu^*(x-\mu)\chi(x-\mu)\} \nonumber \\
{}
\end{eqnarray}
where $x$ labels the points and $\mu$ the forward unit vectors 
in the lattice.
The $U_\mu(x)$ are compact $U(1)$ link variables, $\chi(x)$ are fermion 
fields (Grassmann variables) and $\eta_\mu(x)$ are the Kogut-Susskind phases, 
defined as usual,
\begin{equation}
\eta_\mu(x)=(-1)^{x_1+x_2+...+x_{\mu-1}} \ \ (1\le \mu \le d)
\end{equation}
The fermionic part of the action is a bilinear of the fermion fields,
$S_F=\overline\chi\Delta\chi$, with $\Delta=mI+\Lambda$, and $\Lambda$ 
an antihermitian matrix
\begin{eqnarray}
\Lambda_{x,y}=\frac{1}{2}\sum_\mu\eta_\mu(x)
[\delta_{x+\mu,y}U_\mu(x)-\delta_{x-\mu,y}U_\mu^*(x-\mu)] \\ 
1\le\mu\le d  \nonumber
\end{eqnarray}
Due to its bilinear nature, we can integrate analytically the fermionic 
part of the action in the partition function, thus obtaining
\begin{eqnarray}
Z=\int[d\overline\chi][d\chi][dU] 
\ e^{-\beta S_G -\overline\chi\Delta\chi} \nonumber \\
=\int[dU] \det \Delta \ e^{-\beta S_G} = \int[dU] \ e^{-S_{eff}}
\end{eqnarray}
with the effective action defined as
\begin{equation}
\label{efectiva}
S_{eff}=\beta S_G-\ln{\det\Delta}=\beta S_G-\textrm{tr} \ln\Delta
\end{equation}
Now we are going to consider the loop expansion of the effective action, as
well as the expansion of the determinant of the Dirac operator.

\subsection{The hopping parameter expansion}

Let us examine first the expansion of the effective action \cite{Stamatescu}. 
We have
\begin{eqnarray}
\textrm{tr}\ln\Delta=\textrm{tr}\ln(mI+\Lambda)=
\textrm{tr}\ln(m(I+m^{-1}\Lambda)) \nonumber \\
= V\ln m +\sum_{k\ge1} (-1)^{k+1} \frac{m^{-k}}{k} \textrm{tr}(\Lambda^k)
\label{traza}
\end{eqnarray}
The elements of $\Lambda$ have a natural interpretation as directed links 
on the lattice: due to the local character of the action, only elements 
corresponding to pairs of nearest-neighborhood sites are non zero, so we can 
assign to each non-zero element $\Lambda_{x,y}$ the unique directed link 
which goes from $x$ to $y$. We can thus rewrite the previous traces as
\begin{eqnarray}
\textrm{tr} (\Lambda^k) =
\sum_{x,\mu_1,\mu_2\cdots \mu_k}\Lambda_{x,x+\mu_1}
\Lambda_{x+\mu_1,x+\mu_1+\mu_2}\cdots \nonumber \\
\cdots \Lambda_{x+\mu_1+\cdots+\mu_{k-1},x+\mu_1+\cdots+\mu_{k-1}+\mu_{k}}
\delta_{\mu_1+\cdots+\mu_k,0}
\end{eqnarray}
In this paragraph only, and for the sake of clarity, the $\mu_i$ represent
backward as well as forward unit vectors in the lattice.

We can assign to each non-vanishing term of the trace of 
$\Lambda^k$ a unique closed and oriented loop on the lattice 
constructed with $k$ links. Since loops must be closed, it is obvious that 
$k$ has to be even in order to give a non-zero contribution to (\ref{traza}). 
Consequently we rewrite the trace as
\begin{equation}
\textrm{tr}\ln\Delta=V\ln m -\sum_{k\ge 1}\frac{m^{-2k}}{2k}
\sum_{l\in{\mathcal{G}}_{2k}}\Lambda_l
\end{equation}
where ${\mathcal{G}}_j$ represents all the loops with $j$ links in the class 
considered, and $\Lambda_l$ is a notation for the oriented product of the 
elements of $\Lambda$ around loop $l$.
The net result is real, as for the contribution of
each loop we have the conjugate contribution of the loop taken in the
opposite direction.
To examine more closely this expression, and specifically the
sign of the different terms, we use the definition of $\Lambda$.
For each backward link we get a minus sign.
But for a closed loop, exactly half of the links are backward links, 
so for a loop with $2k$ links we get a factor of $(-1)^k$. 
On the other hand we have to take into account the Kogut-Susskind phases.
One simple way to do this is to absorb them into the links, with
a trivial change of variables, $W_\mu(x)=\eta_\mu(x) U_\mu(x)$.
The corresponding change induced in the plaquettes consists in 
multiplying each of them by $-1$. The final result for the 
effective fermion action expressed in the new variables $W_\mu(x)$ is
\begin{equation}
S^F_{eff} = -V\ln m + \sum_{k\ge 1} (-1)^{k} \frac{m^{-2k}}{2k(2)^{2k}}
\sum_{l\in{\mathcal{G}}_{2k}} W_l
\end{equation}
where $W_l$ means the directed product of the $W$ variables along loop $l$.
We can see from this expression that the effective action is 
frustrated in the sense that there is no single configuration
that minimizes term by term the expansion. Different terms are minimized
by different configurations.
It should be said that even if we have used Kogut-Susskind fermions in our
derivation, the conclusion remains valid for Wilson fermions.

It is interesting to note that the analogous derivation for a
scalar theory coupled to an Abelian gauge field does not show any kind
of frustration. The essential differences with respect to the fermionic case 
are on one hand that the corresponding determinant appears in the denominator,
and on the other hand that in the scalar case the Kogut-Susskind phases are
absent. The final result of the development of the effective action for the 
scalar case is the following:
\begin{equation}
S_{eff}=\beta S_G+V\ln m - \sum_{k\ge 1} \frac{m^{-2k}}{2k(2)^{2k}}
\sum_{l\in{\mathcal{G}}_{2k}} U_l
\end{equation}
In this case each term of the expansion is minimized by the trivial
configuration, so we have no frustration.

\subsection{The expansion of the determinant}

The expansion discussed in the previous section is the most convenient
approach for massive fermions. However one can always proceed directly with 
the expansion of the determinant of $\Delta$, as long as the volume is kept 
finite \cite{MFA1}. So let us consider
\begin{equation}
\label{det}
\det\Delta=\det(mI+\Lambda)=
\sum_{i_1,i_2,\cdots,i_V} \epsilon_{i_1,i_2,\cdots,i_V}
\Delta_{1,i_1}\cdots\Delta_{V,i_V} 
\end{equation}
where $\epsilon_{i_1,i_2,\cdots,i_V}$ is the totally antisymmetric symbol
in $V$ indices. 
Due to gauge invariance and to the properties of the determinant, each 
non-vanishing contribution to (\ref{det}) can be associated to a graph on 
the $L^d$ lattice made up of some number of isolated sites (each one 
contributing with a factor $m$) and some number of closed, directed, simple 
(i.e., no link can appears twice) and non-intersecting loops. In other words, 
after eliminating the isolated sites, all the rest of the lattice should be 
covered with closed loops in such a way that in each site there are exactly 
one incoming and one outcoming link. In particular this implies that the 
number of isolated sites (and so the powers of m) should be even. 

To minimize the contribution of the fermionic part to the effective action
(\ref{efectiva}) one should look for configurations that maximize the 
fermionic determinant. In general, it is impossible to maximize 
simultaneously with a single configuration all the terms in the expansion 
of the determinant (this can be seen easily examining simple 
graphs containing two isolated sites, for example). 

\section{The toy model}
\label{toy-model}

Motivated by the expansions analyzed in the previous section, we have
introduced an Abelian gauge toy model with explicit frustration 
\cite{Edimburgo}. 
We have studied the effect of frustration in this model as a first step 
towards the analysis of the effect of frustration in more realistic models, 
like QED. The model we have studied is a $U(1)$ pure gauge model defined in a 
hypercubic lattice by the following action:

\begin{equation}
S=S_4+S_6=-\beta\sum_{pl}\mathrm{Re}U_{pl}+\beta_6\sum_{p_6}\mathrm{Re}U_{p_6}
\label{action}
\end{equation}
with $\beta , \beta_6 \geq 0$.
The first part ($S_4$) is the standard compact $U(1)$ pure gauge Wilson action.
The second part ($S_6$) is defined as the sum of contributions over all closed 
loops made up with six non-repeated links. 
These loops can be generally classified in three different classes: planar 
loops, loops that involve two planes and loops that involve three planes 
(obviously in the special case $d=2$ only planar loops appear). 

\subsection{Classical ground states and order parameter}
\label{ground}

It is not difficult to realize the existence of frustration in this model
at the classical level in $d> 2$, even if we consider only the $S_6$ 
piece of action (\ref{action}). In fact, it is not possible to minimize 
simultaneously the contributions to the action of all the loops.
One could try to work out the generic minima for some of the classes of
loops but instead, we will describe here,
motivated by our numerical results, a specific class of configurations
that appear to be the classical ground states accessible to the system.
This will permit us to define a natural order parameter for this system.

In the configurations considered the plaquettes take the values $\pm 1$.
It is easy to see that we can find configurations of this type which are
minima for the planar and three-plane loops simultaneously. For the
planar loops, the minimum condition requires that the sign of the
plaquette should be alternate on every plane (for this reason we will call 
these configurations antiferromagnetic, in analogy with magnetic systems). 
Among these configurations we can find a subset that minimize also the 
three-plane loops. There are a total of $2^{(d-1)}$ configurations of 
this type, related by simple symmetry transformations.
Now we should consider the two-plane loops. It is easy to check that these 
loops are not simultaneously minimized in the configurations that we have 
described. In fact, the net contribution to the action of these loops is zero 
in any of the minima considered. In the case $d=2$ only the planar loops appear
in the action; therefore the antiferromagnetic configurations are 
global minima and the system is not frustrated.

On the other hand the minimum for the $S_4$ part of the action is reached for
the homogeneous configuration $U_{pl} = 1$, which becomes another source
of frustration when considering the complete system.

This analysis suggests the introduction of an order parameter for each 
set of parallel planes, the staggered plaquette, defined as follows:
\begin{equation}
P^s_{\mu\nu} =
\frac{1}{V}\sum_x\epsilon(x)\mathrm{Re}U_{\mu\nu}(x), \
\epsilon(x) = (-1)^{x_1+x_2+\cdots+x_4}
\end{equation}
This order parameter is different from zero in the antiferromagnetic 
vacuum and vanishes in the ferromagnetic (homogeneous) one. For the 
configurations described before it takes the values $\pm 1$.

\subsection{The case $d=2$ and the XY model}

We will show in this subsection how, in $d=2$, the thermodynamic limit of 
our model is the XY model.

To this end let us consider first the $S_6$ part of action (\ref{action}). 
In two dimensions it contains only planar loops. It is a well known
fact that in $d=2$ it is possible to replace the links by the elementary 
plaquettes as independent variables of the theory. This is due to the fact that
there are no cubes in the lattice, and hence only one Bianchi identity: the
product of all plaquettes must be equal to $1$. Therefore the number of 
independent plaquettes is $V-1$. If we consider all the plaquettes as
independent variables, the difference in the action is a term of order 
$O(1)$, which is irrelevant in the large $V$ limit.
The plaquette variables can now be assigned to the sites of the dual lattice,
with an action which involves only  nearest neighborhoods.
Now, to obtain the usual form for the XY model, we still need 
another change of variables. We consider the usual bipartition of the planar 
lattice, and we change all the plaquettes of one of the sublattices to minus 
their complex conjugate. If we do this, the final form for the action is
\begin{equation}
S_6=\beta_6 \sum_{<i,j>} \mathrm{Re} (U_i (- U_j^*)) =
- \beta_6 \sum_{<i,j>} \cos (\theta_i-\theta_j)
\end{equation}
where $\sum_{<i,j>}$ means a sum over all pairs of nearest neighborhood sites 
of the (dual) lattice and $\theta_k$ is the angle corresponding to the element
of $U(1)$ in each site. But this is just the action of the XY model on a 
square lattice. As is well known, this model has only one ground state and
does not present frustration.

The contribution of the $S_4$ term to (\ref{action}), expressed in the new 
variables, can be written as
\begin{equation}
S_4=-\beta \sum_i \mathrm{Re} (U_i) = 
-\beta \sum_i \epsilon_i \cos (\theta_i)
\end{equation}
where $\epsilon_i$ equals $\pm 1$ according to which sublattice the site $i$ 
belongs to.
We see that the full action for the two-dimensional model,
\begin{equation}
S=- \beta_6 \sum_{<i,j>} \cos (\theta_i-\theta_j)
-\beta \sum_i \epsilon_i \cos (\theta_i)
\end{equation}
is the action of the X-Y model in the presence of an external staggered
magnetic field. The X-Y model with an external random field has been
extensively studied (\cite{Ma,XYRAND} and references therein) as a simple 
example of frustrated system, 
but we are not aware of studies of the staggered case, so we do not know to 
what extent frustration is relevant for this system.

\section{Numerical simulations}
\label{numeric}

We performed Montecarlo simulations of this model in $d=3$ and $d=4$  
using a standard Metropolis algorithm. 
The observables which were measured in the simulations were the staggered 
plaquette  and the usual normalized plaquette in each plane, and 
the normalized contribution of the 6-loop part of the action, 
\newpage
\begin{eqnarray}
P^s_{\mu\nu} = \frac{1}{V}\sum_x\epsilon(x) \mathrm{Re} U_{\mu\nu} (x) \ ,
\epsilon(x) = (-1)^{x_1+x_2+\cdots+x_d} \\
1\le \mu < \nu \le d \nonumber \\
P_{\mu\nu}=\frac{1}{V}\sum_{x}\mathrm{Re}U_{\mu\nu} \\
P_6=\frac{1}{{\mathcal N} V}\sum_{p_6}\mathrm{Re}U_{p_6}
\end{eqnarray}
where $\mathcal N$ means the number of 6-loops per volume.
We also measured for completeness the imaginary part of these 
quantities, that gave results consistent with zero in all our 
simulations.

In order to obtain a qualitative understanding of the phase diagram of this
model, our strategy was to explore the parameter space over several lines 
of constant $\beta$. To locate the possible transition points we did a 
number of annealing cycles over $\beta_6$, while maintaining $\beta$ fixed to 
a given value. The most extensive simulations were done in $d=4$, and we
discuss this case in more detail below. 

\begin{figure}[h]
\psrotatefirst
\psfig{figure=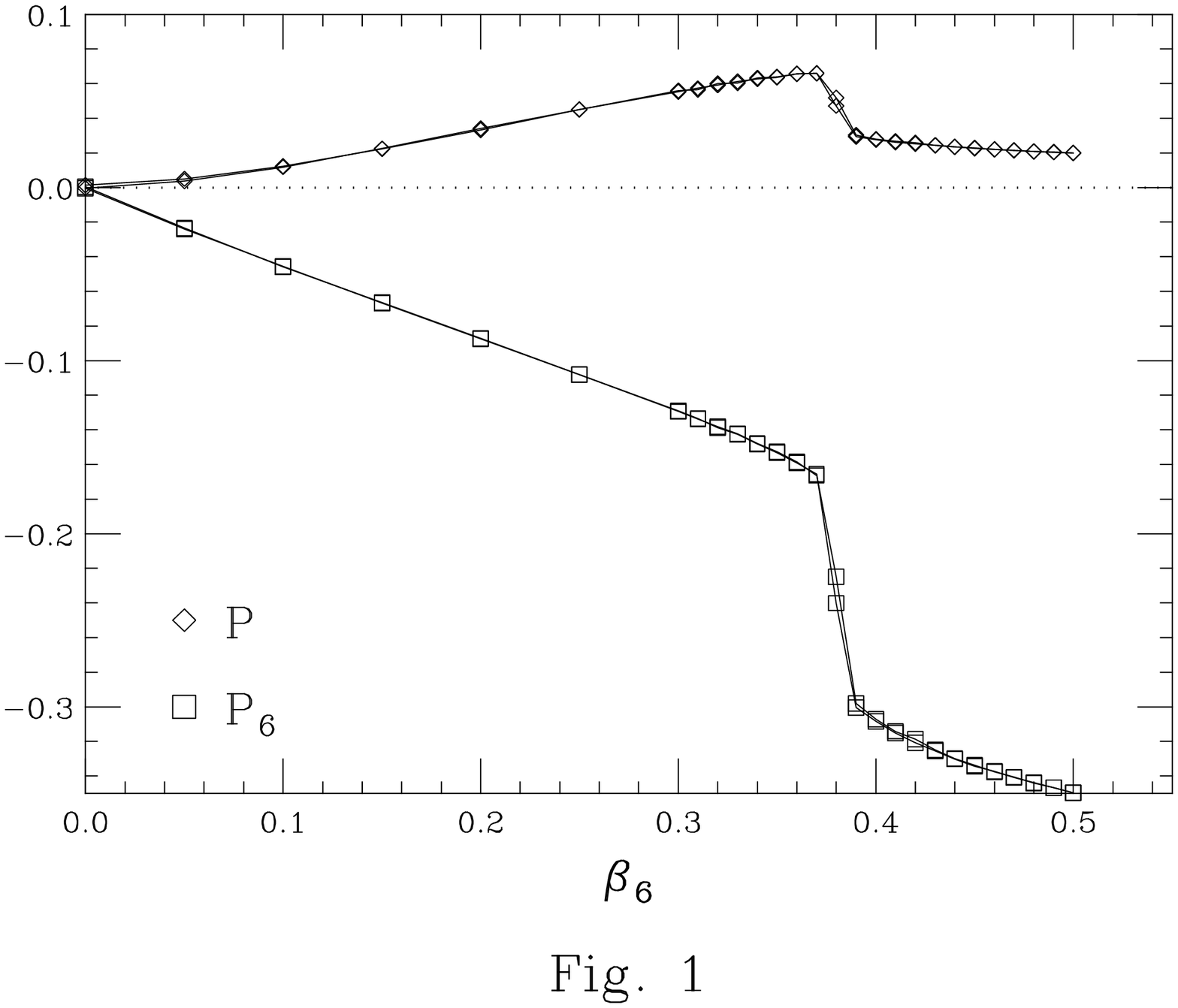,angle=90,width=270pt}
\caption{Plaquette and 6-loop hysteresis cycles at $\beta=0$ against $\beta_6$}
\label{6d3_e}
\end{figure}

\begin{figure}[h]
\psrotatefirst
\psfig{figure=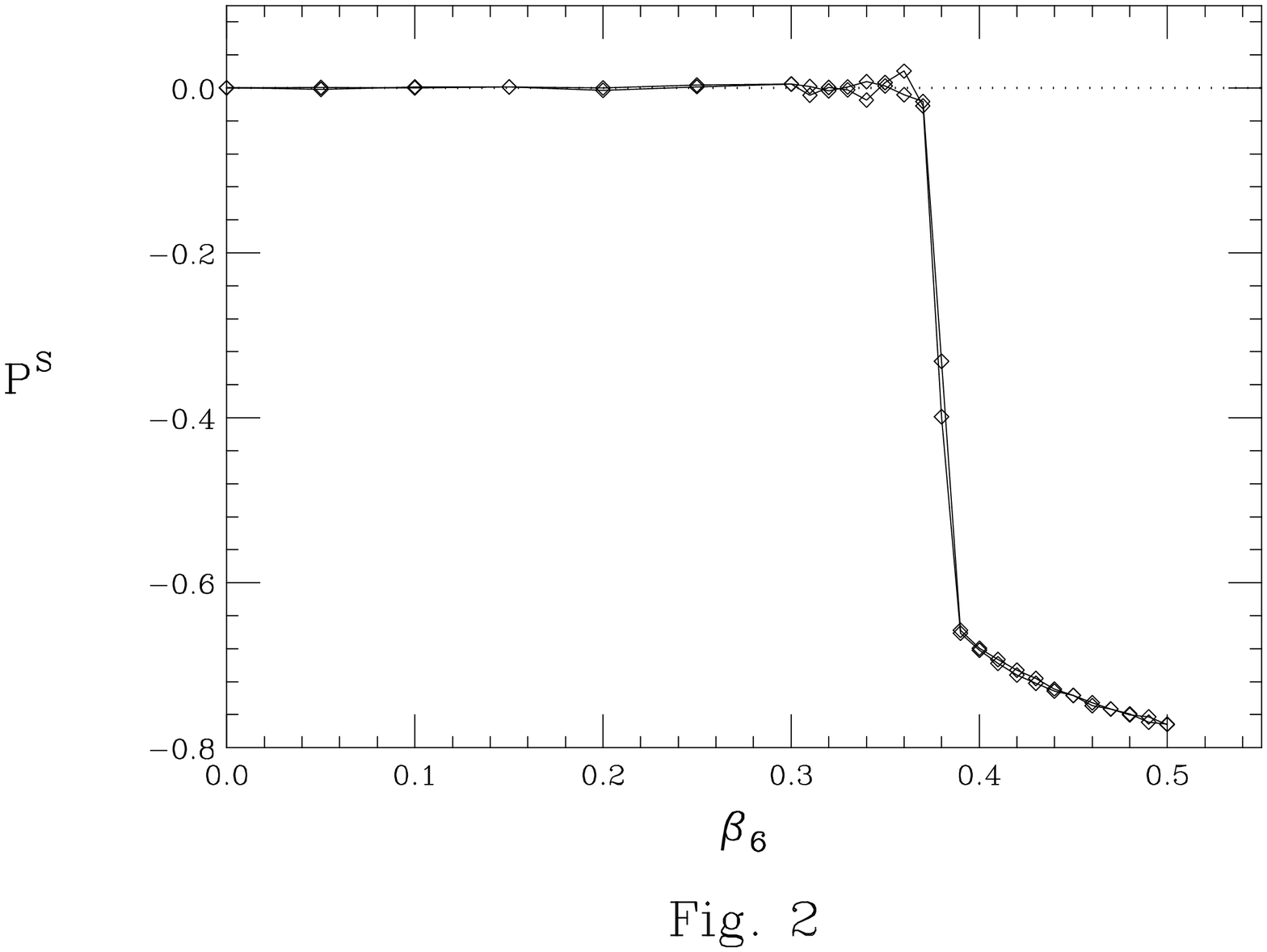,angle=90,width=270pt}
\caption{Staggered Plaquette hysteresis cycle at $\beta=0$ against $\beta_6$}
\label{6d3_ma}
\end{figure}

As we have shown before, in $d=2$ our model is equivalent to the X-Y model.
It is well known that this model presents a continuous transition at a value
of the coupling constant $\beta_6\simeq 0.9$, at least for $\beta=0$.

In Fig.\ \ref{6d3_e} and \ref{6d3_ma} we show the results for the
annealing cycles in $d=3$ at $\beta=0$. The figures suggest the existence of 
a transition into a phase with a non-vanishing value of the staggered 
plaquette at a smaller value of the coupling constant than in $d=2$,  
$\beta_6\simeq 0.38$. However, a more careful study and larger
statistics are necessary to establish
unambiguously the order of the transition. 

\subsection{The system in $d=4$}

\begin{figure}[h]
\psrotatefirst
\psfig{figure=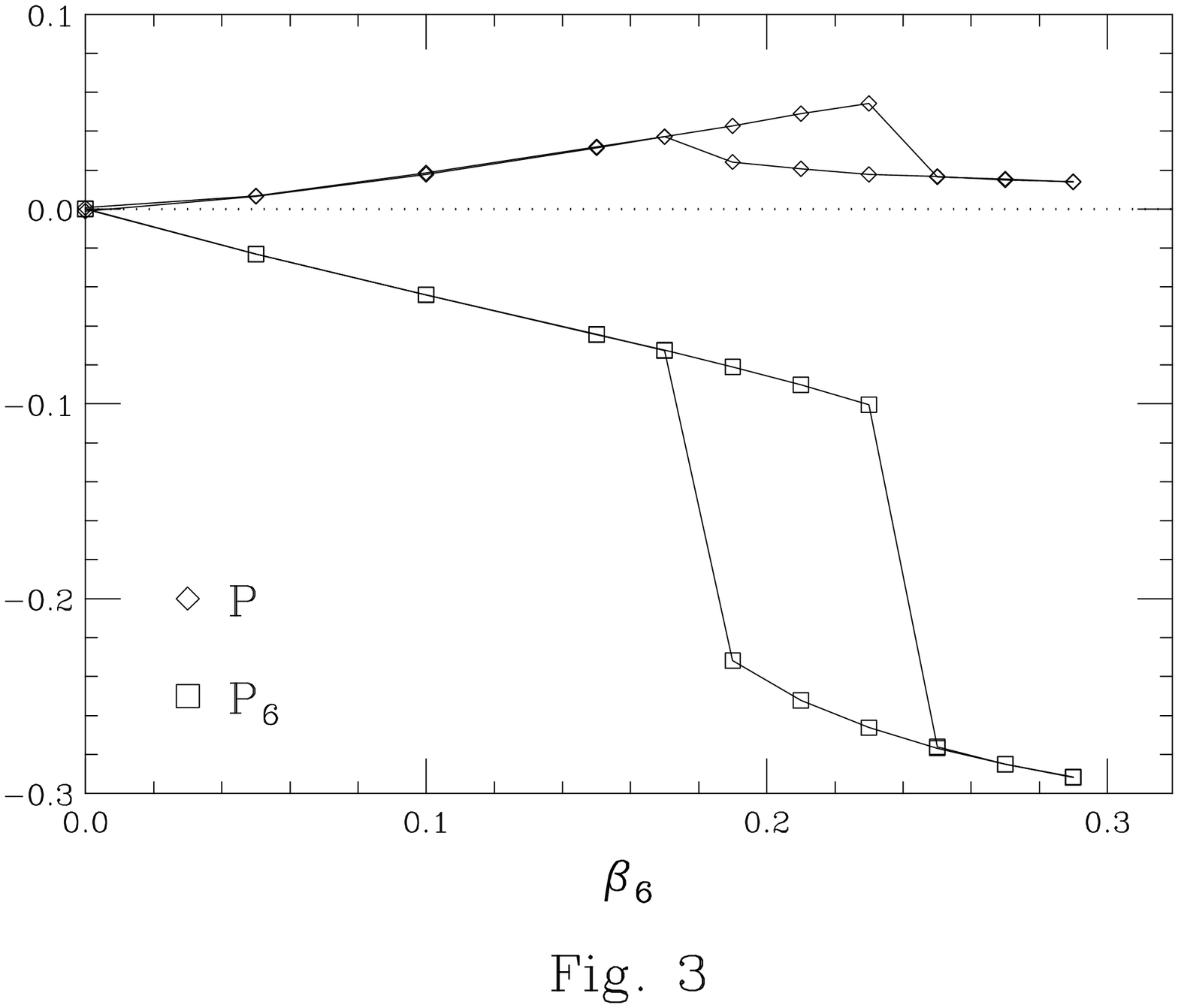,angle=90,width=270pt}
\caption{Plaquette and 6-loop hysteresis cycles at $\beta=0$ against $\beta_6$}
\label{6d4_e}
\end{figure}

\begin{figure}[h]
\psrotatefirst
\psfig{figure=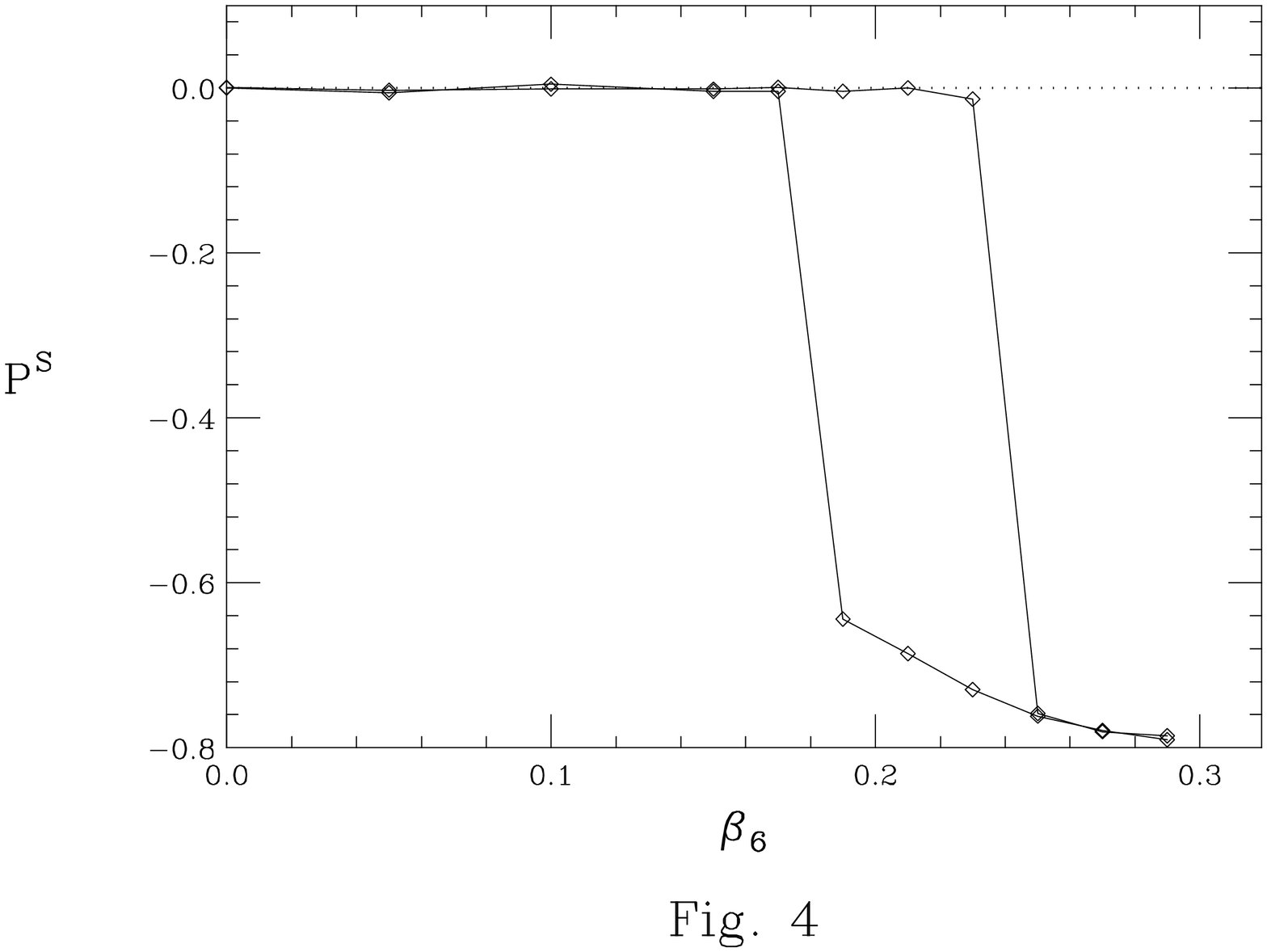,angle=90,width=270pt}
\caption{Staggered Plaquette hysteresis cycle at $\beta=0$ against $\beta_6$}
\label{6d4_ma}
\end{figure}

We simulated the four-dimensional system along several lines of constant 
$\beta \ge 0$. First, we did several runs at different
$\beta$, while maintaining
$\beta_6$ fixed to zero, in order to obtain properly thermalized configurations
that we used later to start the annealing cycles in $\beta_6$. 
For each value of $\beta$, we started the annealing with the thermalized 
configuration obtained in this way. 
After we finished the simulation for a given value of $\beta_6$, we changed
it by some small amount and repeated the process starting with the last 
configuration generated at the previous value of $\beta_6$. Proceeding in 
this way we did a complete forward-backward cycle.
These simulations were done in $4^4$ and $6^4$ lattices.

We show in Fig.\ \ref{6d4_e} and \ref{6d4_ma} the resulting annealing 
cycles for some of the magnitudes we have measured at $\beta=0$.The 
staggered plaquette is shown here only for one of the planes (see the 
discussion below).
We see a very clear hysteresis effect signaling a strong first-order
phase transition. In order to check that this is not the effect of a
bad thermalization around a continuous transition we compared cycles
obtained by increasing the simulation time over two orders of magnitude,
from $1000$ to $100000$ Montecarlo sweeps, and no change in the width of the
hysteresis region was seen. Also no significant finite-size effects were
found in our simulations. 
We also implemented an over-relaxation procedure which was combined with
the Metropolis algorithm; again, no change in the measured magnitudes was 
observed.
Concerning the nature of the large $\beta_6$ phase, we can see in figure 
4 that the staggered plaquette shows a sudden increase from
zero, when entering into this phase. If we compare its value for the
different planes we see that it only differs in sign, with the signs
being consistent with the ones obtained from the analysis of the classical
ground state configurations done in section (\ref{ground}). The simulations 
at larger values of $\beta_6$, produced configurations which were essentially 
the classical ground states described before.

\begin{figure}[h]
\psrotatefirst
\psfig{figure=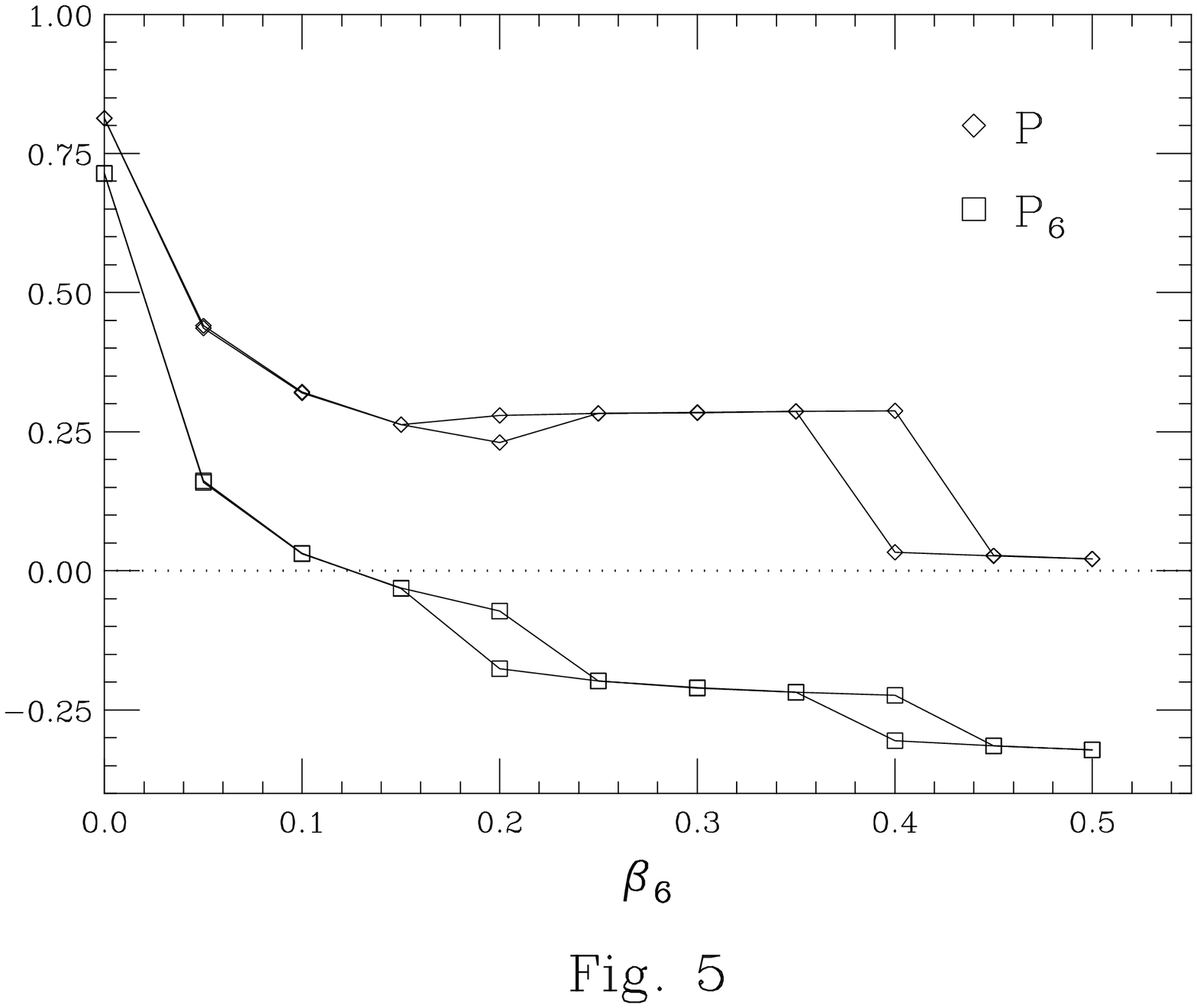,angle=90,width=270pt}
\caption{Plaquette and 6-loop hysteresis cycles 
at $\beta=1.5$ against $\beta_6$}
\label{6d4c4_e}
\end{figure}

\begin{figure}[h]
\psrotatefirst
\psfig{figure=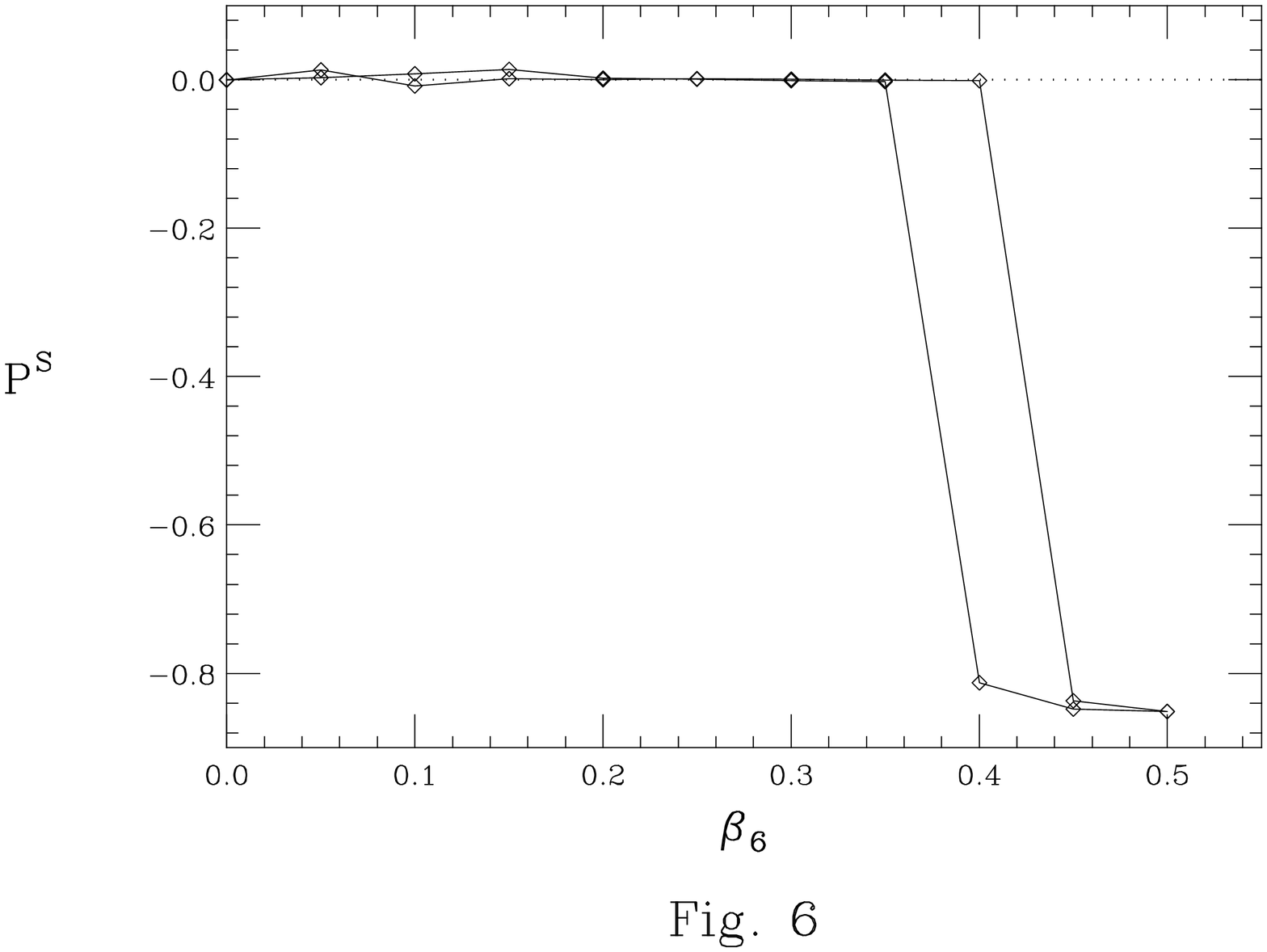,angle=90,width=270pt}
\caption{Staggered Plaquette hysteresis cycle at $\beta=1.5$ against $\beta_6$}
\label{6d4c4_ma}
\end{figure}

\begin{figure}[h]
\psrotatefirst
\psfig{figure=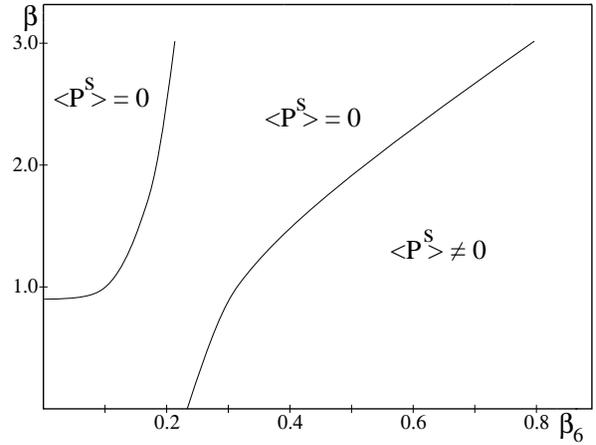,angle=270,width=220pt}
\caption{Phase diagram}
\label{phase_diagram}
\end{figure}

Now let us discuss our results for the lines at constant $\beta>0$.
For small values of $\beta$, the results of the simulations are qualitatively
the same as for $\beta=0$, apart from the fact that the transition point 
occurs at higher values of $\beta_6$ for higher values of $\beta$. 
But we know that this system, for $\beta_6=0$, has a first-order
transition at $\beta\simeq 1.0$, the usual confined-deconfined
transition for the standard compact $U(1)$ pure gauge theory 
\cite{Jersak,U1_compacto,Isabel}.
So it is very likely that a line of transition points emerges 
from the $\beta$ axis and, indeed, this is what we found. We did simulations
at several valued of $\beta$ above the transition point. 
In Fig.\ \ref{6d4c4_e} and \ref{6d4c4_ma} we show the results obtained 
by cycling over $\beta_6$ while keeping 
$\beta=1.5$. We can clearly see here two transitions: 
the first one should correspond to the continuation of the usual 
confined-deconfined transition (the staggered plaquette remains zero for 
this transition), whereas the second one corresponds to the transition 
to the antiferromagnetic phase, as clearly shown by the jump 
of the staggered plaquette. The results for all the $\beta$ values are
consistent with this picture, and permit us to obtain a qualitative phase
diagram, that we show in Fig.\ \ref{phase_diagram}. 
We found three phases separated by two first-order lines.
The antiferromagnetic phase is characterized by a non-zero value
of the staggered plaquette, which is zero in the other two phases. In this
phase we have several states, to be precise eight states, related by 
a spontaneously broken translational symmetry. The other two phases are the 
continuation of the confined and unconfined phases of the standard compact 
pure gauge model.

\section{Comments}
\label{comments}

Our starting point was the observation that competing interactions are present
in the effective gauge theory obtained, after integration on
Grassmann fields, from a gauge-fermion theory on the lattice, such as
$QED$. Inspired from analogies with condensed matter models like spin-glasses
in which frustration and disorder conspire to give non standard phases to 
the system, we asked if the rich, and in some sense unexpected fenomenology 
observed in $3d$ and $4d$ lattice $QED$ can be ascribed to the presence of 
non standard vacua.

The first step in our analysis has been the study of a simpler model, in 
which frustration is introduced by hand, but as suggested by the hopping 
parameter expansion. In the $d=3$ and $d=4$ cases we have seen
clear signals of the existence of an antiferromagnetic phase in which 
translational invariance is broken. This is not totally surprising since the
solutions of classical equations of motion, at least in a limiting
case of the parameter space, share this property.
A crucial point in the analysis has been the availability of a 
clear and simple order parameter, inspired from the properties of
these classical configurations. The $d=2$ case was found analytically
equivalent to the $X-Y$ model with staggered external magnetic field
in the thermodynamical limit.

The simple model we have considered
has only a finite number of competing interactions, so on general grounds we
cannot expect to find spin glass-like phases. On the contrary, in the
full model, the number of frustrated terms in the effective action
is diverging with the volume of the system. We do not know if this is
enough to simulate in some way
the inclusion of disorder, that in a spin glass system
is the other ingredient for the appearance of non standard phases: this is an
issue for future work. The first problem we have to face is the
definition of a suitable order parameter.

The list of features in three and four dimensional $QED$ that lack
an explanation is long, and we suspect that, at least some of them,
could be related to frustration at the level of the effective action.

\acknowledgments

This work has been partly supported through a CICYT (Spain) - INFN (Italy)
collaboration.

E. Follana acknowledges support from Caja de Ahorros de La Inmaculada.

\end{document}